\documentstyle[12pt,epsfig,rotate]{article}

\def\GeV{{\mathrm GeV}}
\newcommand\Dqns{\Delta q^{\rm ns}}

\newcommand{\id}{\mbox{d}}
\def\beq{\begin{equation}}
\def\eeq#1{\label{#1}\end{equation}}
\def\eeqn{\end{equation}}

\def\beqa{\begin{eqnarray}}
\def\eeqa#1{\label{#1}\end{eqnarray}}
\def\eeqan{\end{eqnarray}}

\def\overbar#1{\overline{#1}}

\let\bar=\overbar

\def\Dslash{\not{\hbox{\kern-4pt $D$}}}
\def\dslash{\not{\hbox{\kern-2pt $\del$}}}

\def\msb{{\bar{\ssstyle M \kern -1pt S}}}

\def\Title#1{\begin{center} {\Large {\bf #1} } \end{center}}

\begin{document}

\Title{The Spin Structure of the Nucleon}

\bigskip\bigskip

%+\addtocontents{toc}{{\it A.B. Author}}
%+\label{authorStart}

\begin{raggedright}  

{\it G.K. Mallot\index{Mallot, G.K.}\\
EP Division, CERN\\
CH-1211 Geneva, Switzerland
}
\bigskip\bigskip
\end{raggedright}

%% Please take out the next three lines for the proceedings
%%
\vbox to 0pt{\vspace{18cm}\hrule \vspace{1mm}\em \noindent 
Invited talk presented on the XIX International Symposium on 
Lepton and Photon Interactions, August 9--14, 1999, Stanford University, CA}

\section{Introduction}

The discovery \cite{EMC_88a, EMC_89a} by the European Muon Collaboration
(EMC) that the axial charge of the proton is much smaller than expected from
the Ellis--Jaffe sum rule \cite{ElJ74} implies that the nucleon spin is not simply made
up by the quark spins.
This surprising result was confirmed by a series of experiments: the Spin Muon Collaboration 
experiment at CERN \cite{SMC_98a,SMC_99a,SMC_97a}, the SLAC experiments E142 \cite{E142_96a},
E143 \cite{E143_98a}, E154 \cite{E154_97a, E154_97c}, E155 \cite{E155_99a, Mitchell_99} and 
the Hermes experiment \cite{Hermes_97a,Hermes_98a, Hermes_99a} at DESY. 
They increased the kinematic range and the precision of the data to a level where 
QCD analyses begin to become a powerful tool like in the unpolarised case.

Albeit the immense theoretical progress and the wealth of data the original problem 
remains that the origin of the nucleon's spin is not yet understood.
Apart from the quark spins, $\Delta\Sigma=\Delta u + \Delta d + \Delta s$,
the gluon spin, $\Delta g$, and orbital angular momentum, $L$, 
must play a major r\^ole in making up the nucleon spin
\begin{equation}
\label{eq:heliSR}
\frac{1}{2}=\frac{1}{2}\Delta\Sigma+L_q+\Delta g + L_g.
\end{equation}
It is well known from both, experiment and theory, that at high $Q^2$ about
half of the nucleon's longitudinal momentum is carried by the gluons.
The same sharing was predicted for the total angular momentum of the
nucleon \cite{JiT96a}.
In the Quark Parton Model the polarised quark distribution functions,
\begin{equation}
\Delta q = \left(q^{+}-q^{-}\right) +
           \left(\overline{q}^{+}-\overline{q}^{-}\right),
\end{equation}
are related to the spin-dependent structure function $g_1$ by
\begin{equation}
g_1(x,Q^2) = \frac{1}{2}\sum_{f} e_f^2 \Delta q_f(x,Q^2),
\end{equation}
where $f$ runs over the quark flavours and $e_f$ are the electrical
quark charges. The notations $q^{+(-)}$ refer to
parallel (antiparallel) orientation of the quark and nucleon spins.

Experimentally the spin-dependent structure functions, $g_1$ and $g_2$,
are obtained from the measured event-number asymmetries,
$A^{\rm raw}_\parallel$, for longitudinal orientation of the target and
lepton spins, and $A^{\rm raw}_\perp$ for transverse target polarisations.
These raw asymmetries range for the proton typically from
a few per cent at large $x$ to a few parts per thousand at small $x$.
In the lepton-nucleon asymmetries, $A_\parallel$ and $A_\perp$,
the uncertainties of the raw asymmetries get amplified by the factor
$1/P_bP_tf$ accounting for the incomplete beam and target polarisations,
$P_b$ and $P_t$, and the dilution factor, $f$.
Typical target materials contain a large fraction of unpolarisable nucleons
and $f$ denotes the fraction of the total spin-averaged cross section arising
from the polarisable nucleons.
For the target materials used, $f$ varies from about 0.13 (butanol),
over 0.17 (ammonia) to 0.3 ($^3$He). For deuterated butanol and ammonia
$f$ is 0.23 and 0.3, respectively, while for the proton and deuteron gas
targets $f$ is close to unity.
The neutron structure functions are either obtained from the combination of
proton and deuteron data or from experiments using $^3$He targets.
The deuteron asymmetries are
slightly reduced from the average of proton and neutron asymmetries due to the
D-state component in the deuteron wave function.
The $^3$He asymmetry is mainly due to the unpaired neutron,
however a small proton contribution has to be corrected for.
Due to the cancellation of the isotriplet part in $g_1^{\rm d}$
the measurements using deuteron targets are most sensitive to the
flavour-singlet part and thus to the Ellis--Jaffe sum rule.
The structure functions, $g_1$ and $g_2$, are related to the virtual photon
asymmetries,  $A_1$ and $A_2$, via
\begin{eqnarray}
\label{eq:ApAt}
A_\parallel=&D(A_1+\eta A_2),  \hskip 1cm & A_\perp=d(A_2-\xi A_1),\\
\label{eq:A1A2}
A_1=&\displaystyle\frac{g_1-\gamma^2 g_2}{F_1},  \hskip  1cm & A_2=\gamma\frac{g_1+g_2}{F_1}.
\end{eqnarray}
Here $F_1=F_2(1+\gamma^2)/2x(1+R)$
is the well-known spin-averaged structure function and
$\gamma^2=Q^2/\nu^2$, $\eta$, and $\xi$ are kinematic factors, which
are small in most of the kinematic domain covered by the data.
The variables $\nu$ and $Q^2=-q^2$ denote respectively the energy transfer
and negative square of the 4-momentum transfer.
The kinematic factors, $D$ and $d$, account for the incomplete
transverse polarisation of the virtual photon.
With longitudinal target polarisation predominantly $g_1$ is determined, while
experiments with transverse target polarisation are sensitive to $g_1+g_2$.
The virtual photon asymmetries are bounded by $|A_1|\le 1$ and $|A_2|\le\sqrt{R}$,
where $R=\sigma_L/\sigma_T$ is the longitudinal-to-transverse photoabsorption
cross-section ratio known, like $F_2$, from unpolarised deep inelastic scattering.

\begin{figure}[htb]
\begin{center}
\epsfig{width=0.50\hsize,
%bbllx=47, bblly=143, bburx=560, bbury=668,
file=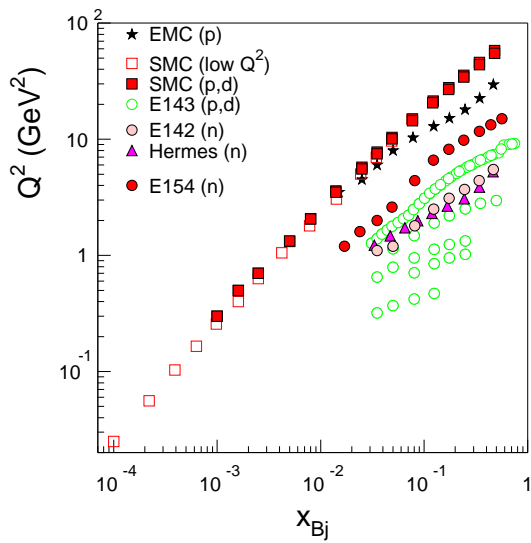}
\caption{Kinematic ranges of the individual experiments. The double logarithmic scale
emphasises the new data in the small-$x$, small-$Q^2$ region from a special
SMC trigger.}
\label{fig:kine}
\end{center}
\end{figure}

\begin{table}[htb]
\begin{center}
\caption{Parameters of the experiments}
\label{tab:exp}
\begin{tabular}{llrrcl@{$\,\le\,$}c@{$\,\le\,$}lc}
\hline
\hline
Experiment  &Lab.  &\multicolumn{2}{c}{Beam}& $\langle Q^2\rangle$&
 \multicolumn{3}{c}{$x$ range}& Targets\\
            &      &&                       &     [GeV$^2$]         &
\multicolumn{3}{c}{} &\\
\hline
E-80/130 &S\sc lac &  22~GeV&e    & ~~4 & 0.18  &$ x $& 0.7& p \\
EMC      &C\sc ern & 200~GeV&$\mu$&  10 & 0.01  &$ x $& 0.7& p \\
SMC      &C\sc ern & 190~GeV&$\mu$&  10 & 0.003 &$ x $& 0.7& p, d \\
E-142    &S\sc lac &  29~GeV&e    & ~~3 & 0.03  &$ x $& 0.8& $^3$He\\
E-143    &S\sc lac &  29~GeV&e    & ~~3 & 0.03  &$ x $& 0.8& p, d \\
E-154    &S\sc lac &  48~GeV&e    & ~~5 & 0.014 &$ x $& 0.7& $^3$He\\
E-155    &S\sc lac &  48~GeV&e    & ~~5 & 0.01  &$ x $& 0.9& p, d ($^6$LiD)\\
Hermes   &D\sc esy &  27~GeV&e    & ~~3 & 0.023 &$ x $& 0.6& $^3$He, p, d \\
\hline
\hline
\end{tabular}
\end{center}
\end{table}

\section{Experiments and status of structure function data}
The experimental approach of the three recent series of experiments
on the structure function $g_1$ is rather different. The highest momentum 
transfers, $Q^2$, and the lowest values of $x$-Bjorken were reached by the 
SMC experiments at the 190~GeV CERN muon beam. The rather low intensity of muon 
beams required thick solid-state polarised targets. The two target cells allow 
a simultaneous measurement of targets with opposite polarisations \cite{SMC_99b}. 
Both, butanol (p,d) and ammonia (p) targets were used.
The to date most precise data come from the SLAC experiments E154 and E155,
which followed the earlier E142 and E143 experiments. They were performed
at the high-intensity, 49~GeV SLAC electron beam with a thin 
$^3$He gas target (E154) and solid state NH$_3$ and LiD targets (E155) \cite{E155_99b}.
Due to the possible rapid change in beam polarisation only one target is 
needed. Several magnetic spectrometers determined the momentum of the 
scattered electrons.
The advantage of the internal gas target of Hermes \cite{Hermes_98b} in the 
Hera electron storage ring is its low mass which enables this experiment to obtain
precise semi-inclusive data, where in addition to the scattered lepton
additional hadrons are detected. Another advantage is the absence of 
unpolarised nucleons in the hydrogen and deuterium gas targets.
Two spin rotators before and after the detector provide parallel or
antiparallel longitudinal polarisation of the electrons at the 
interaction point. The polarisation of the gas target can be inverted
within milliseconds.
The rather low energy of the Hera electron beam of 27~GeV makes 
the interpretation of the data not always straight forward.
The kinematic coverage of these experiments shown in Fig.~\ref{fig:kine}
reflects the different energies of the incident leptons.

\begin{figure}[htb]
\begin{center}
\epsfig{width=0.48\hsize, bbllx=47, bblly=143, bburx=500, bbury=668,
file=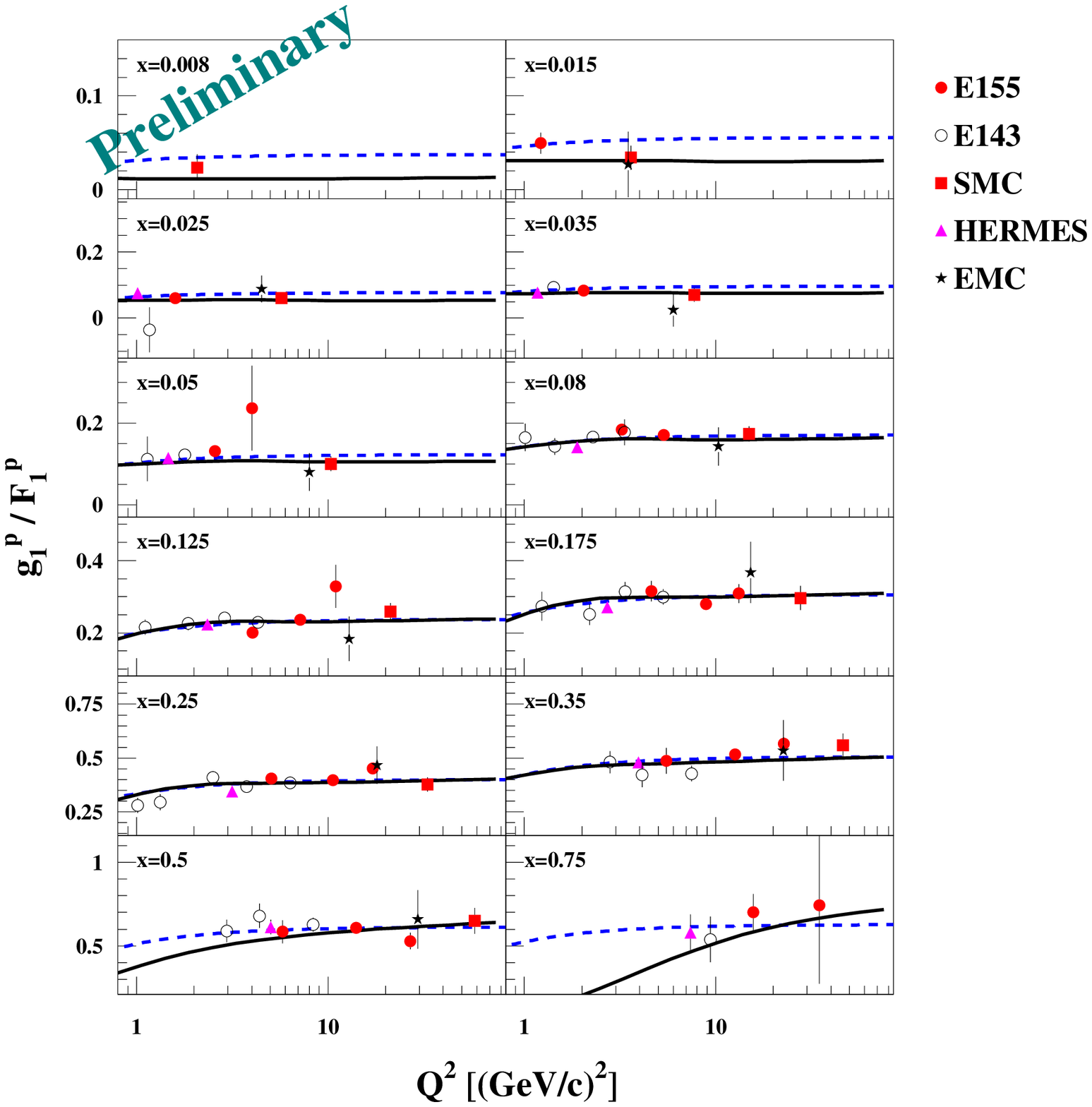}
\hfill
\epsfig{width=0.48\hsize, bbllx=47, bblly=143, bburx=500, bbury=668,
file=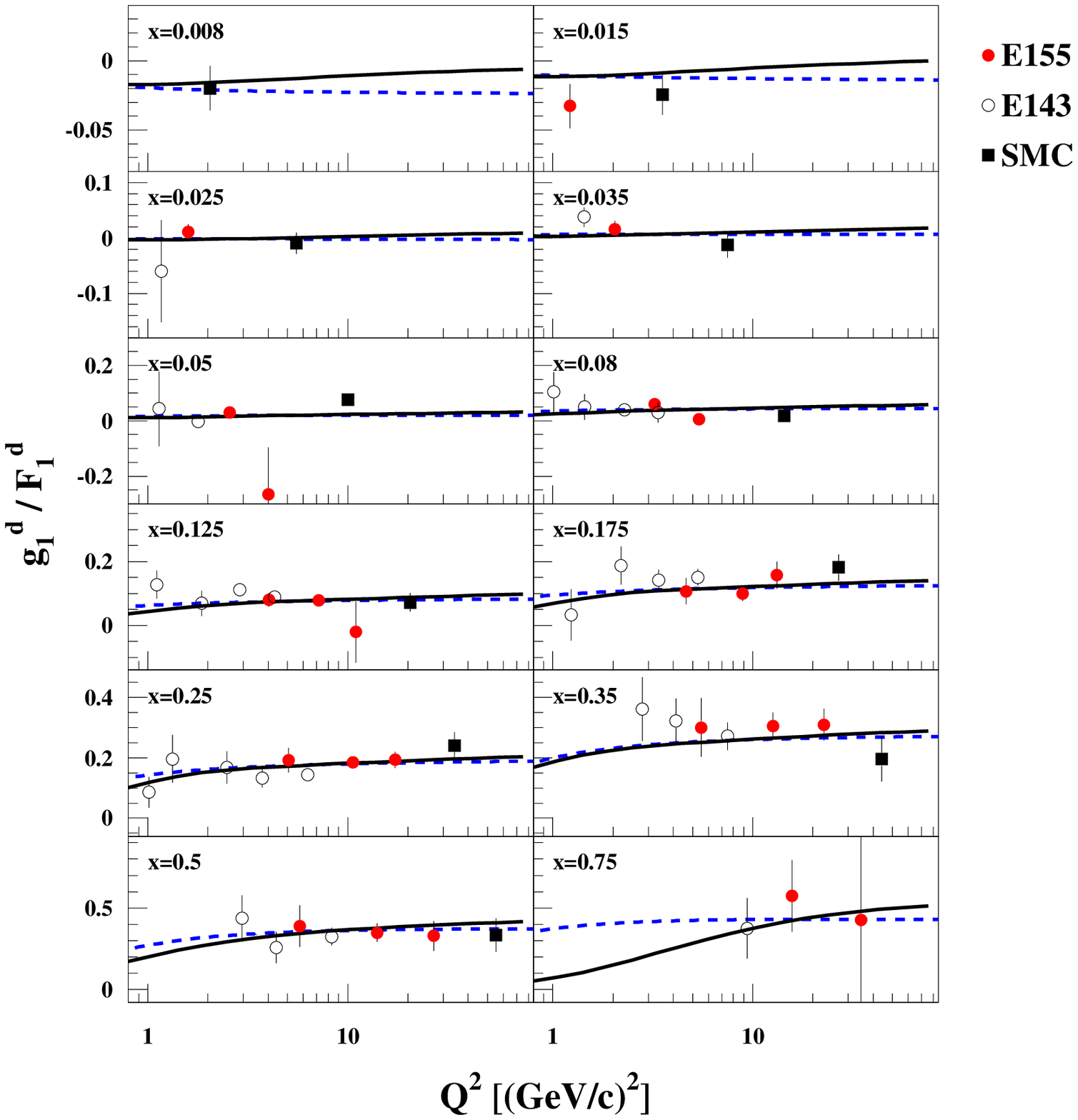}
\caption{Proton (left) and deuteron structure functions $g_1(x,Q^2)$ 
for $Q^2>1~\GeV^2$. The proton data from E155 are preliminary. Also shown 
are a phenomenological fit to the data (dashed) and the E154 QCD-fit 
(solid) \cite{E154_97b}}.
\label{fig:g1_xq}
\end{center}
\end{figure}

\begin{figure}[htb]
\begin{minipage}[t]{0.45\hsize}
\begin{center}
\epsfig{width=\hsize,
file=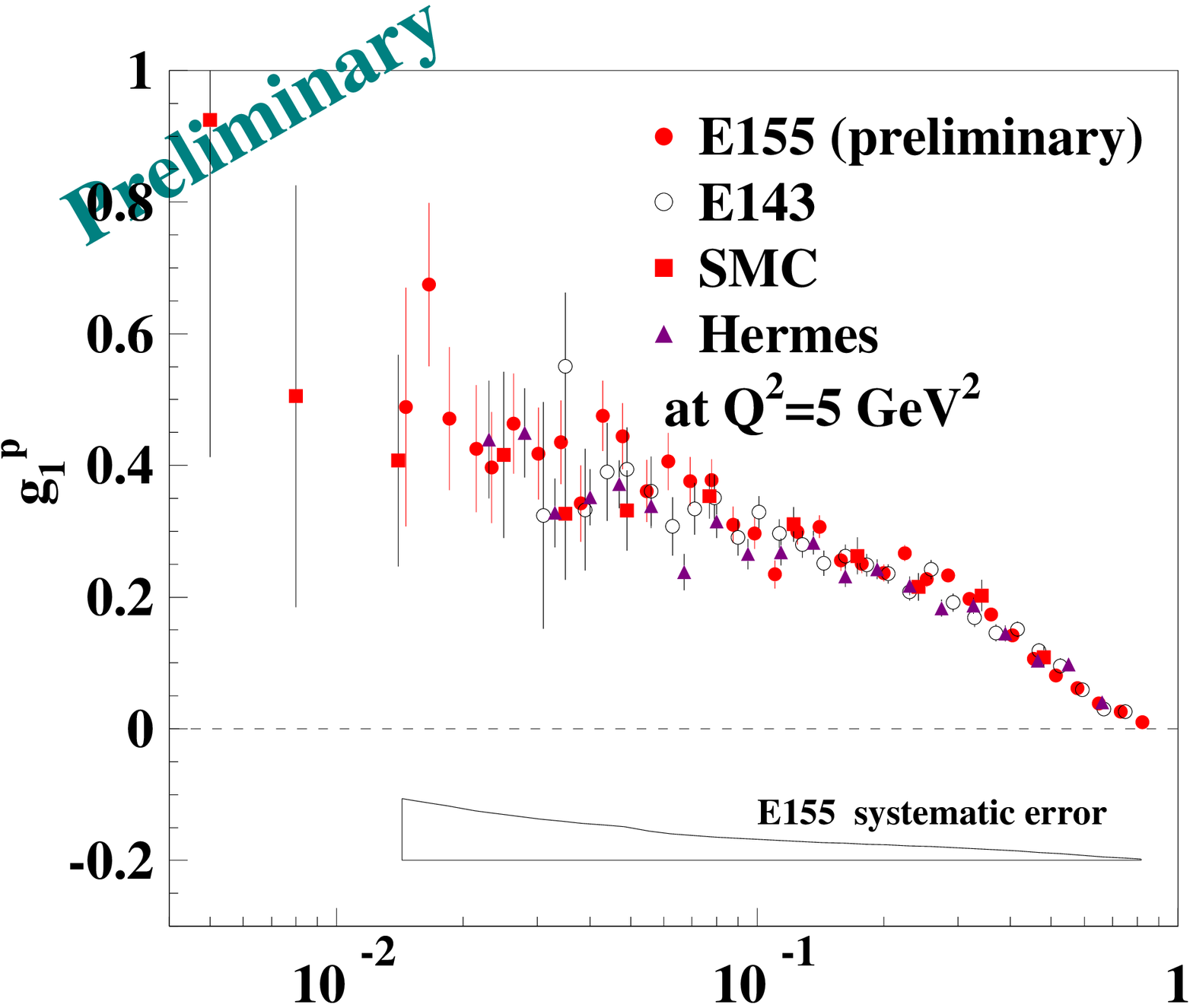}

\epsfig{width=\hsize,
file=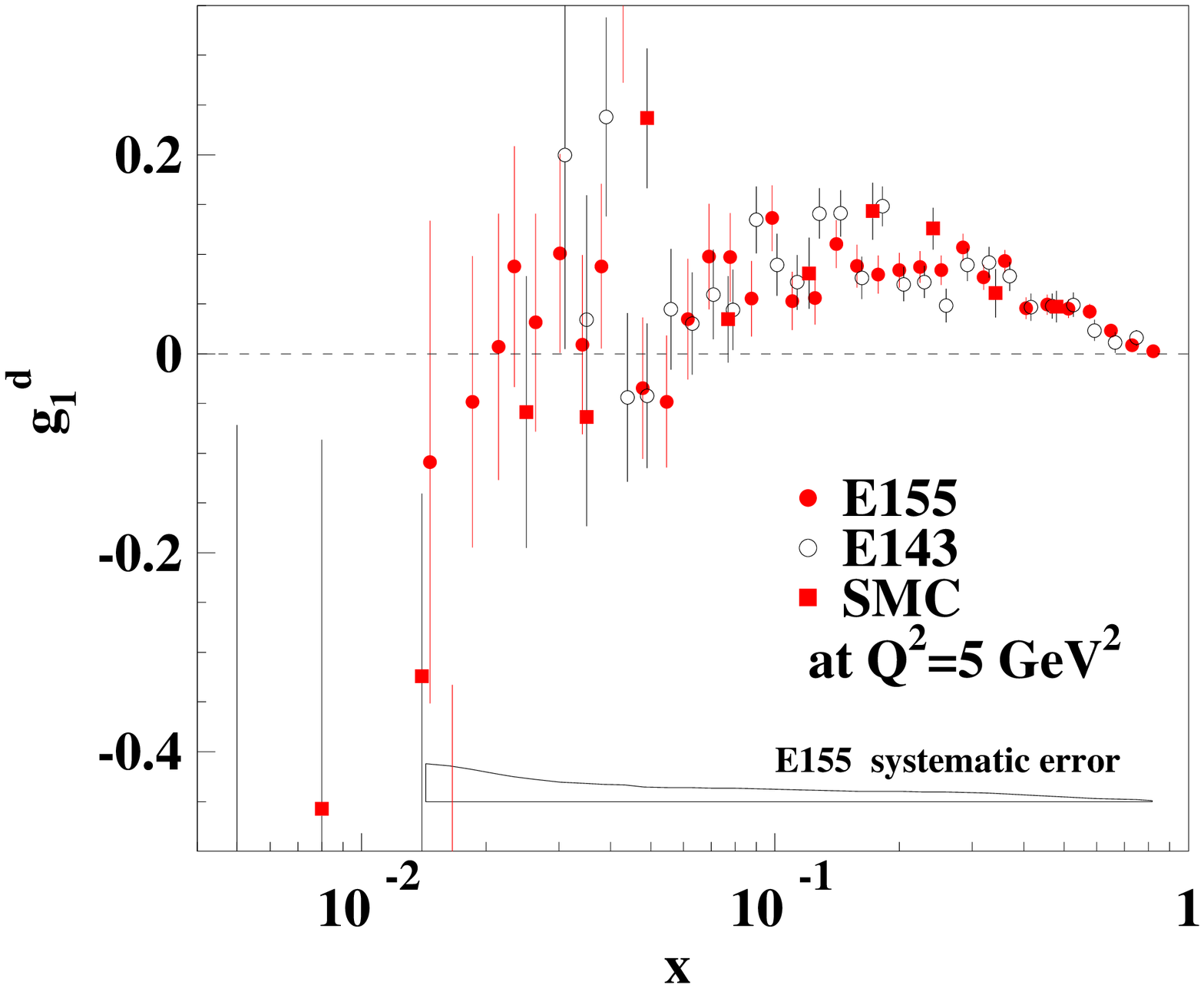}
\end{center}
\end{minipage}
\hfill
\begin{minipage}[t]{0.45\hsize}
\begin{center}
\epsfig{width=\hsize,
file=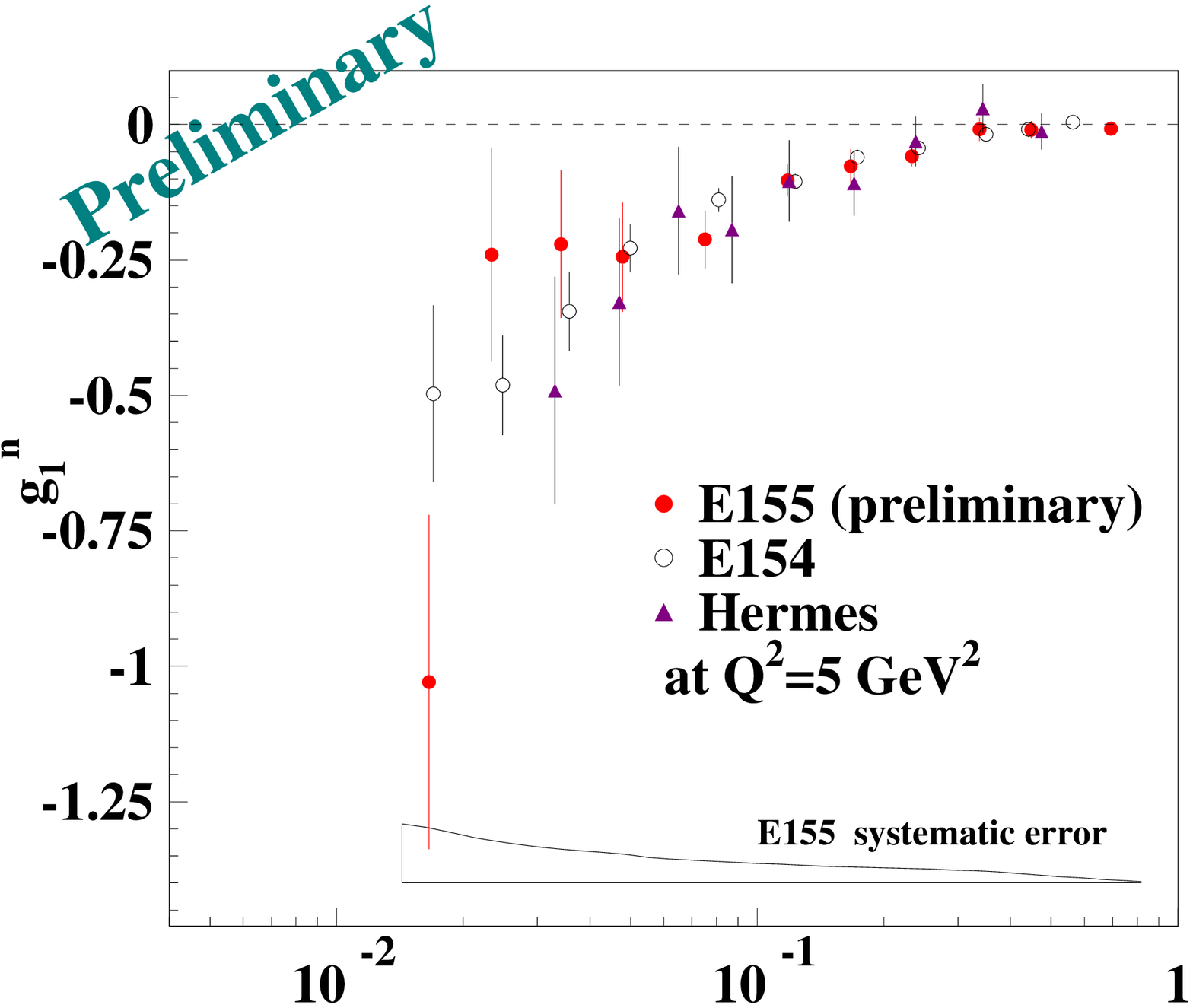}
\caption{The structure function $g_1(x,5~\GeV^2)$ for the proton (left, top),
the deuteron (left, bottom) and the neutron (top) \protect\cite{E155_web}. 
The E155 proton data are preliminary.  Also shown is the systematic error of the
E155 data.}
\label{fig:g1}
\end{center}
\end{minipage}
\end{figure}

The asymmetries $g_1/F_1(x,Q^2)$ as compiled by the 
E155 collaboration are shown for $Q^2>1~\GeV^2$ in 
Fig.~\ref{fig:g1_xq} \cite{E155_99a, E155_web}. 
For the proton an indication of a positive $Q^2$-dependence is 
visible in the small-$Q^2$ region, while the deuteron asymmetries show
no such trend.
In order to compare the data taken at different momentum transfers,
the structure function data, $g_1(x,Q^2)$, were evolved to a common
value of $Q^2=5~\GeV^2$. The results for the proton, the deuteron and the
neutron are shown in
Figs.~\ref{fig:g1}.
All data sets are in good agreement. While for the proton there is
(not yet) an indication of a decreasing $g_1$ towards small $x$ as
expected from QCD evolution, this effect is clearly visible for the
neutron.

\begin{figure}[htbp]
\begin{center}
\begin{minipage}[t]{0.35\hsize}
\begin{center}
\epsfig{width=\hsize,
file=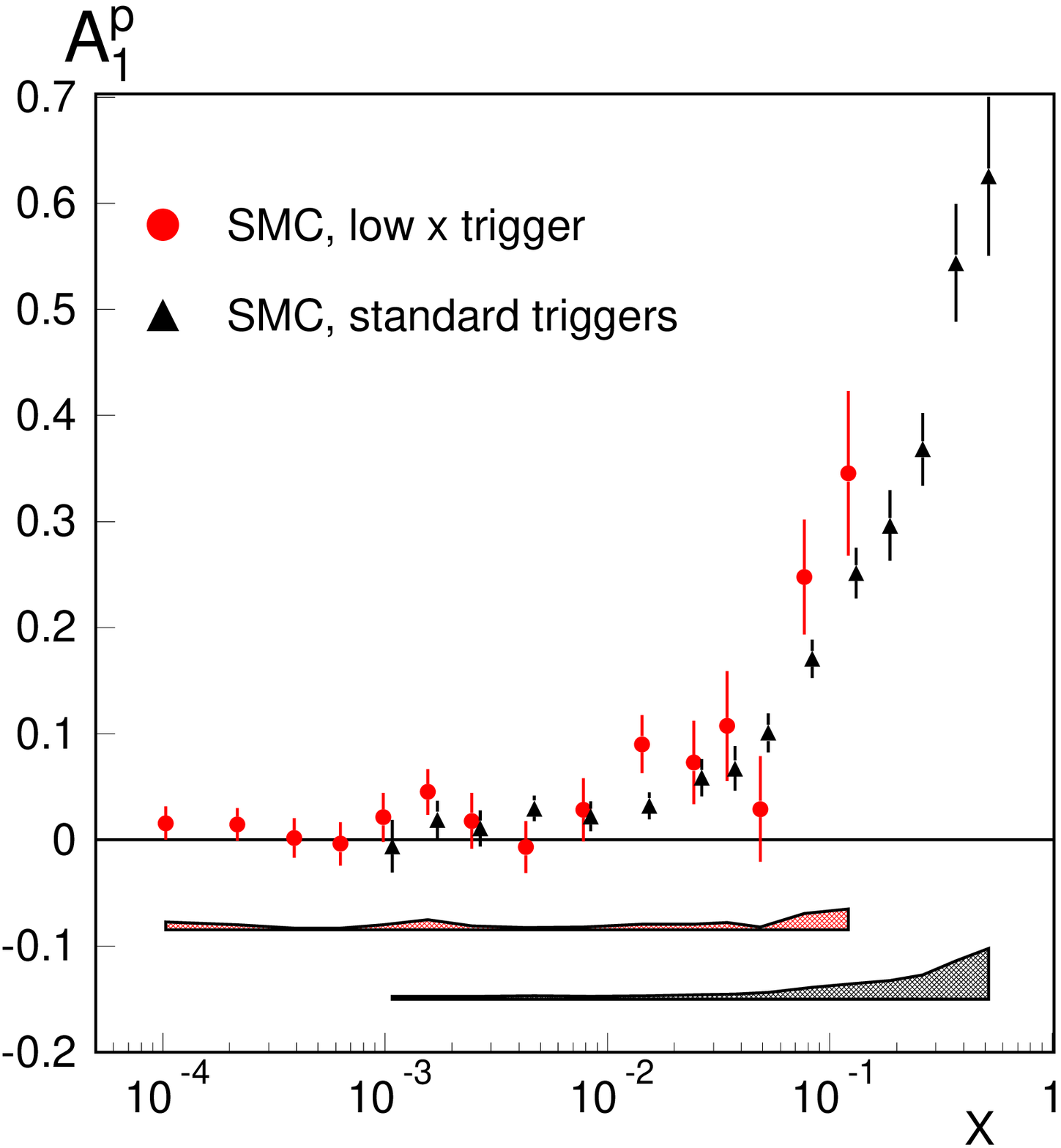}
\end{center}
\end{minipage}
\hskip 1cm
\begin{minipage}[t]{0.35\hsize}
\begin{center}
\epsfig{width=\hsize,
file=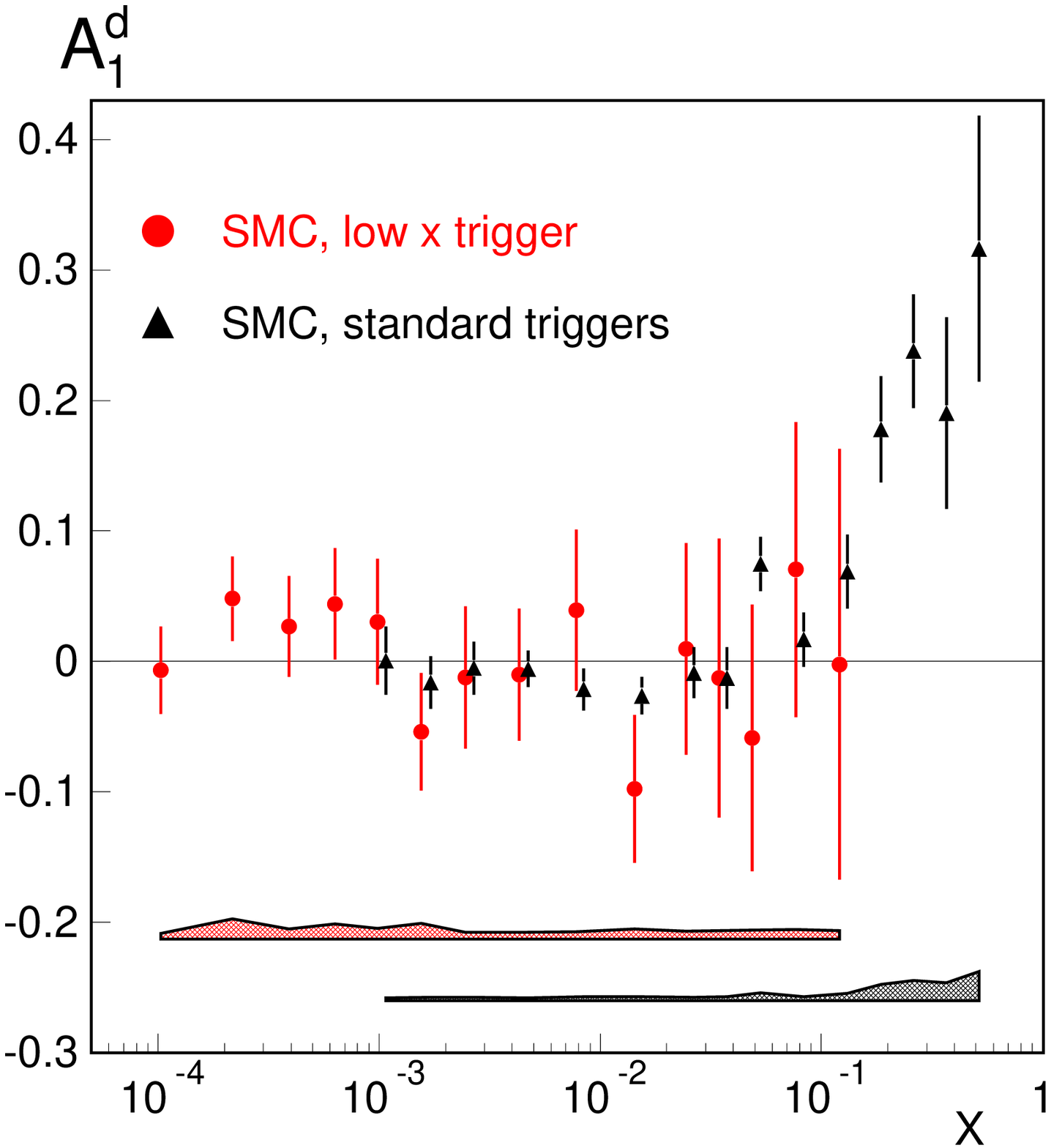}
\end{center}
\end{minipage}
\caption{
Asymmetries, $A_1(x)$, for the proton (left) and the deuteron (right) from the SMC low-$x$
and standard triggers \protect\cite{SMC_99a}. The small-$x$ trigger extends to $x=6\cdot10^{-5}$.}
\label{fig:SMC_asy}
\end{center}

\bigskip
\begin{center}
\begin{minipage}[t]{0.35\hsize}
\begin{center}
\epsfig{width=\hsize,
file=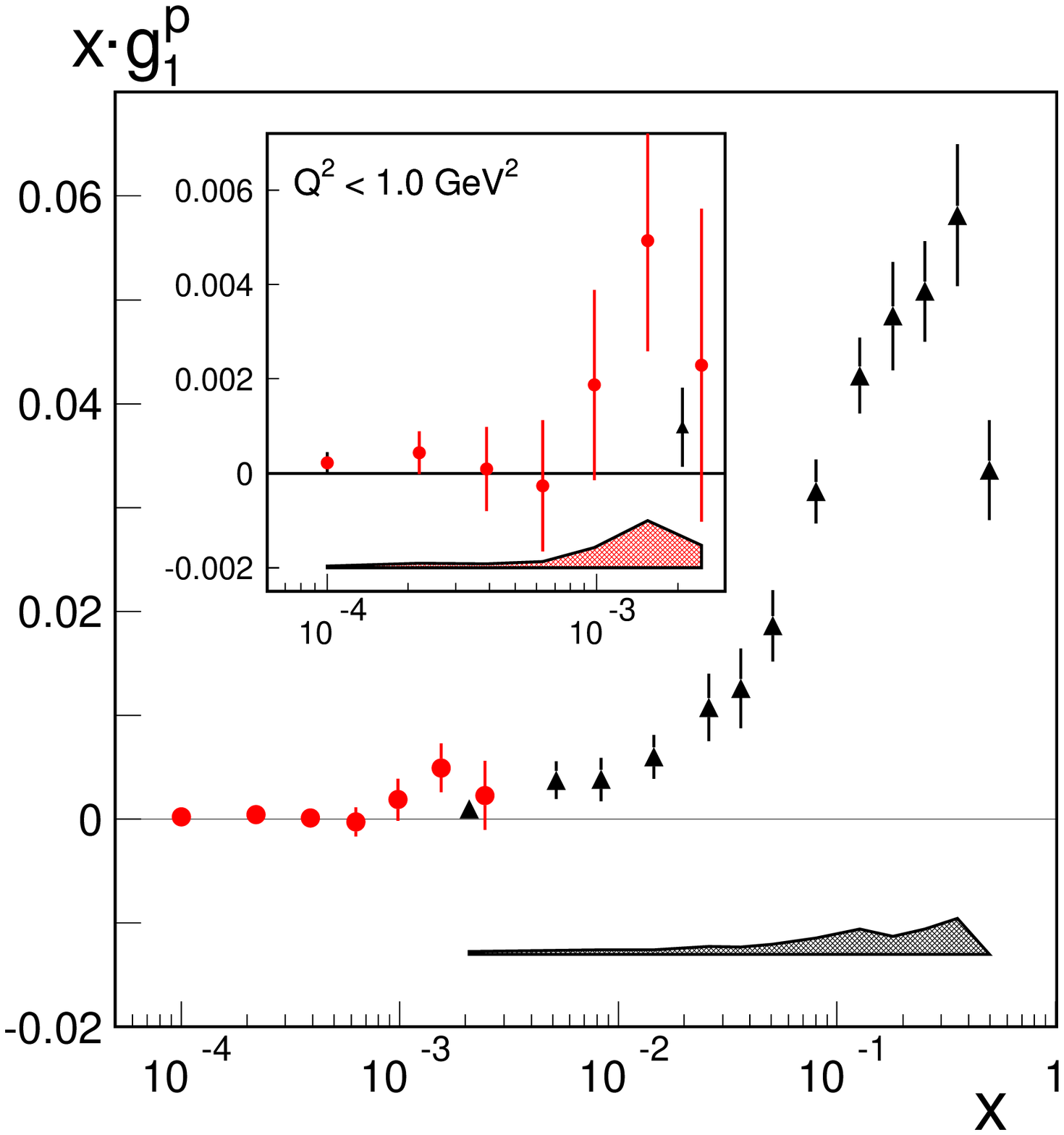}
\end{center}
\end{minipage}
\hskip 1cm
\begin{minipage}[t]{0.35\hsize}
\begin{center}
\epsfig{width=\hsize,
file=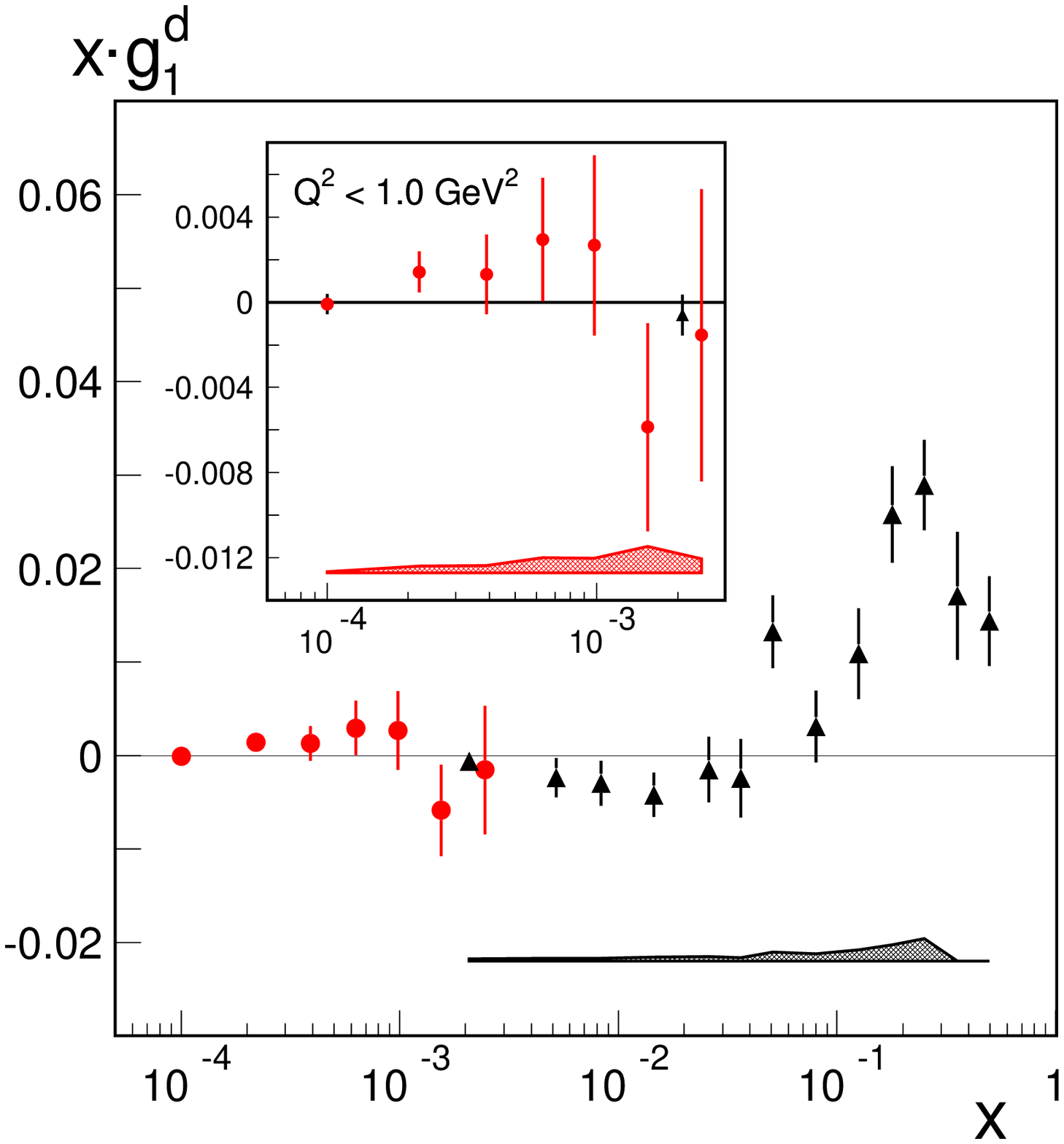}
\end{center}
\end{minipage}
\hfill
\caption{Structure functions, $x\cdot g_1(x)$, for the proton (left) and the deuteron (right)
from the SMC \protect\cite{SMC_99a}. The value of $Q^2$ is below 1~GeV$^2$  for $x<0.003$.
}
\label{fig:SMC_xg1}
\end{center}
\end{figure}

Recently the SMC has published data down to $x=6\cdot10^{-5}$, where
the corresponding momentum transfer is only $Q^2=0.01~\GeV^2$ \cite{SMC_99a}.
These data were taken with a special low-$x$ trigger. The large radiative
corrections at low $x$ were suppressed by requiring the presence of an
additional hadron. This requirement also reduces the contamination
by events from muon-electron scattering, which appear at
$x=m_{\mathrm e}/m_\mu$. Additional kinematic cuts were applied to further
reduce this background.
The asymmetries and structure functions taken with this special 
low-$x$ trigger are shown together with those from the standard physics
triggers in Figs.~\ref{fig:SMC_asy} and \ref{fig:SMC_xg1}. The 
structure functions in the new low-$x$ region do not exhibit any unexpected
behaviour.

\begin{figure}[htb]
\begin{center}
\epsfig{width=0.5\hsize,file=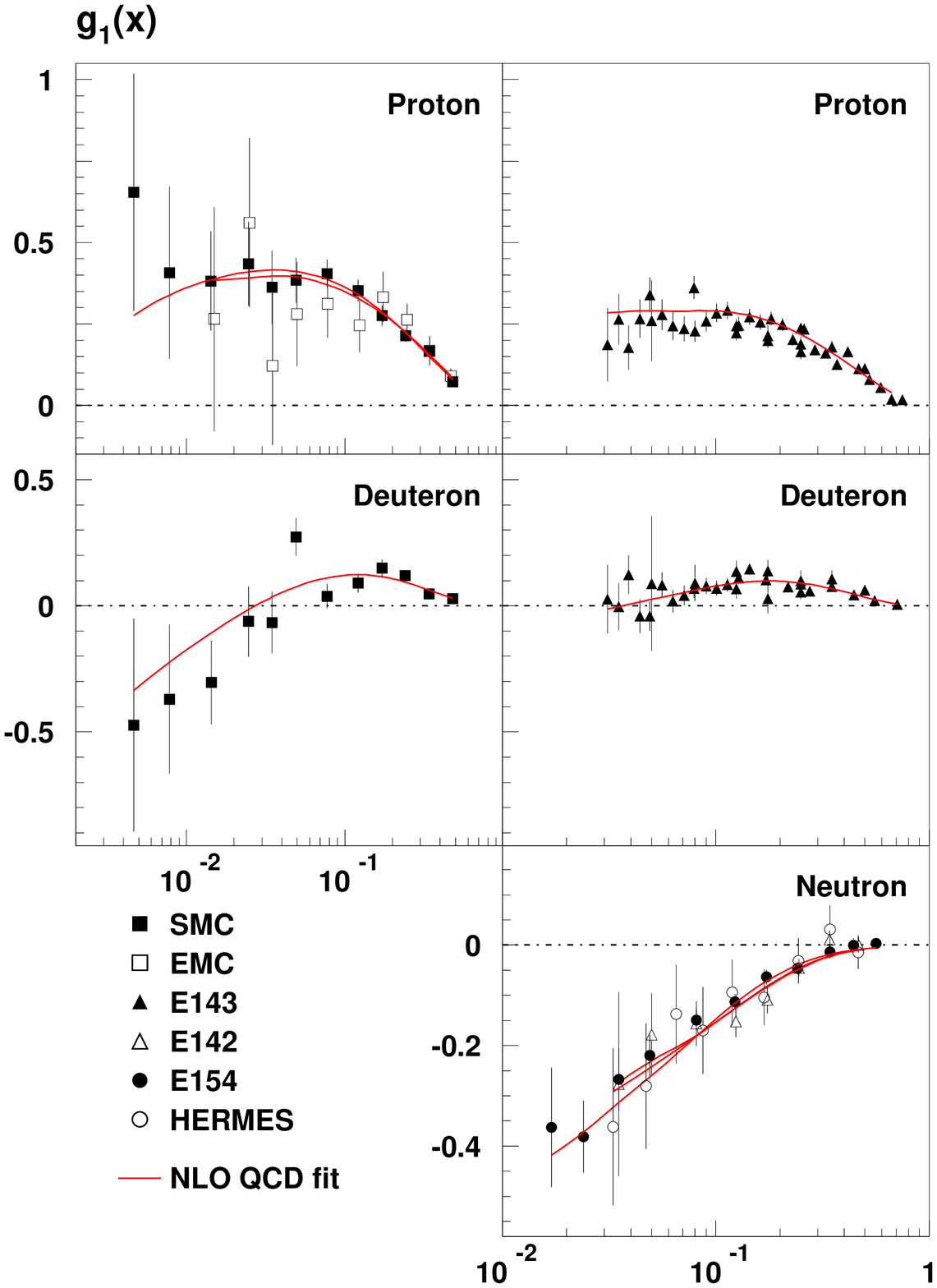}
\caption{SMC next-to-leading order QCD fit to $g_1$ data of the
proton, the deuteron and the neutron. The curves represent the
fit for the different values of $Q^2$ of the data points
\protect\cite{SMC_98b}.}
\label{fig:SMC_qcd}
\end{center}
\end{figure}

\section{QCD analyses of $g_1$ data}

As in the unpolarised case the $Q^2$ evolution of the structure functions
is predicted by QCD. With the increasing precision of the data and after
the next-to-leading order splitting functions were calculated 
\cite{MeN_96a,Vog_96a} QCD analyses
have become a powerful tool in the understanding of polarised structure
function and parton distribution functions. 

The evolution equations are expressed in terms of the polarised splitting
functions, $\Delta P$, and the singlet and non-singlet quark distribution
functions, $\Delta \Sigma$ and $\Dqns$ 
\begin{eqnarray}
\label{eq:glap}
\frac{\id }{\id t}\Dqns&=&\frac{\alpha_s(t)}{2\pi}\Delta
                              P_{qq}^{\rm ns}\otimes\Dqns,\\
\nonumber\\
\label{eq:glapsi}
\frac{\id}{\id t}
       \left(
          \begin{array}{c}
             \Delta\Sigma\\
             \Delta g\\
            \end{array}
          \right)
       &=&\frac{\alpha_s(t)}{2\pi}
       \left(
          \begin{array}{cc}
             \Delta P^{\rm s}_{qq} & 2n_f \Delta P^{\rm s}_{qg}\\
             \Delta P^{\rm s}_{gq} &  \Delta P^{\rm s}_{gg}
             \end{array}
          \right)
       \otimes
       \left(
           \begin{array}{c}
              \Delta\Sigma\\
              \Delta g
              \end{array}
           \right),
\end{eqnarray}
with $t=\ln Q^2/\Lambda^2$ and the number of active flavours, $n_f$.
The evolution of $\Delta \Sigma$ mixes with that of the polarised gluon 
distribution function, $\Delta g$.
The structure function $g_1$ is then given by

\begin{equation}
g_1^{\mathrm p,n} = \frac{1}{12}\left( \pm \Delta q_3 + \frac{1}{3}\Delta q_8\right)
\otimes C^{\mathrm ns}+
 \frac{1}{9}\Bigl(\Delta\Sigma\otimes C^{\mathrm s} + \Delta g \otimes 2 n_f C^{\mathrm g}\Bigr)
\end{equation}
with the Wilson coefficients, $C$. The plus and minus symbols refer to the proton and the neutron,
respectively.  The non-singlet parton distribution functions are given by
\begin{eqnarray}
\Delta q_3(x,Q^2) &=& \Delta u(x,Q^2) - \Delta d(x,Q^2),\\
\Delta q_8(x,Q^2) &=& \Delta u(x,Q^2) + \Delta d(x,Q^2) - 2 \Delta s(x,Q^2).
\end{eqnarray}
The QCD fit is then performed in a particular renormalisation and factorisation scheme using 
parametrisations for $\Delta\Sigma$, $\Dqns$ and $\Delta g$ at an initial value of $Q^2=Q_i^2$.
The fit can either be performed in $x$ or in moment space ($n$ space).

\begin{figure}[htb]
\begin{minipage}[t]{0.40\hsize}
\begin{center}
\epsfig{width=\hsize,file=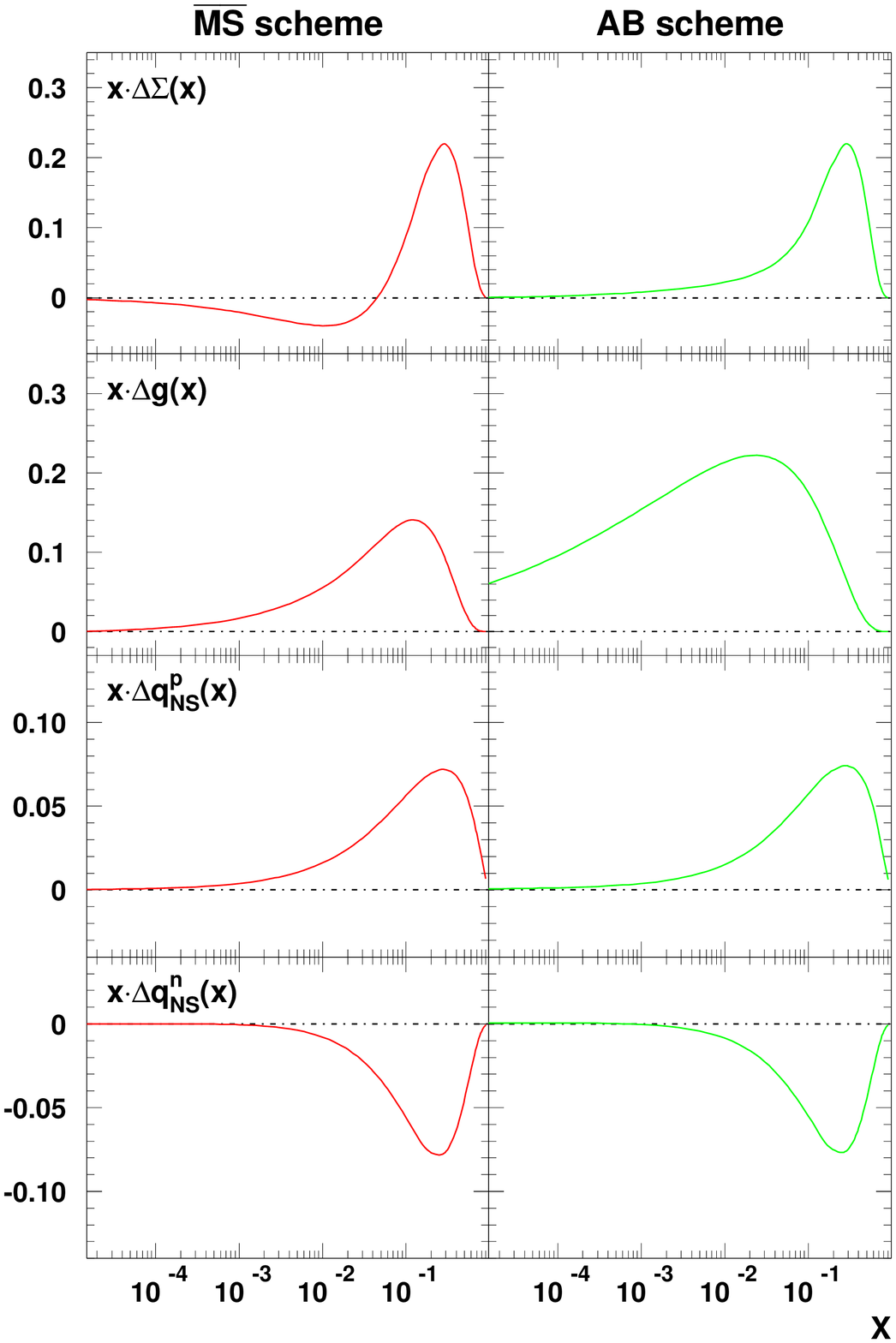}
\caption{SMC next-to-leading order QCD fit, comparison
of the fits in the $\overline{\mathrm MS}$ and AB schemes.
The distributions are shown at $Q^2=5~\GeV^2$
\protect\cite{SMC_98b}.}
\label{fig:SMC_pdf}
\end{center}
\end{minipage}
\hfill
\begin{minipage}[t]{0.50\hsize}
\begin{center}
\epsfig{width=\hsize,file=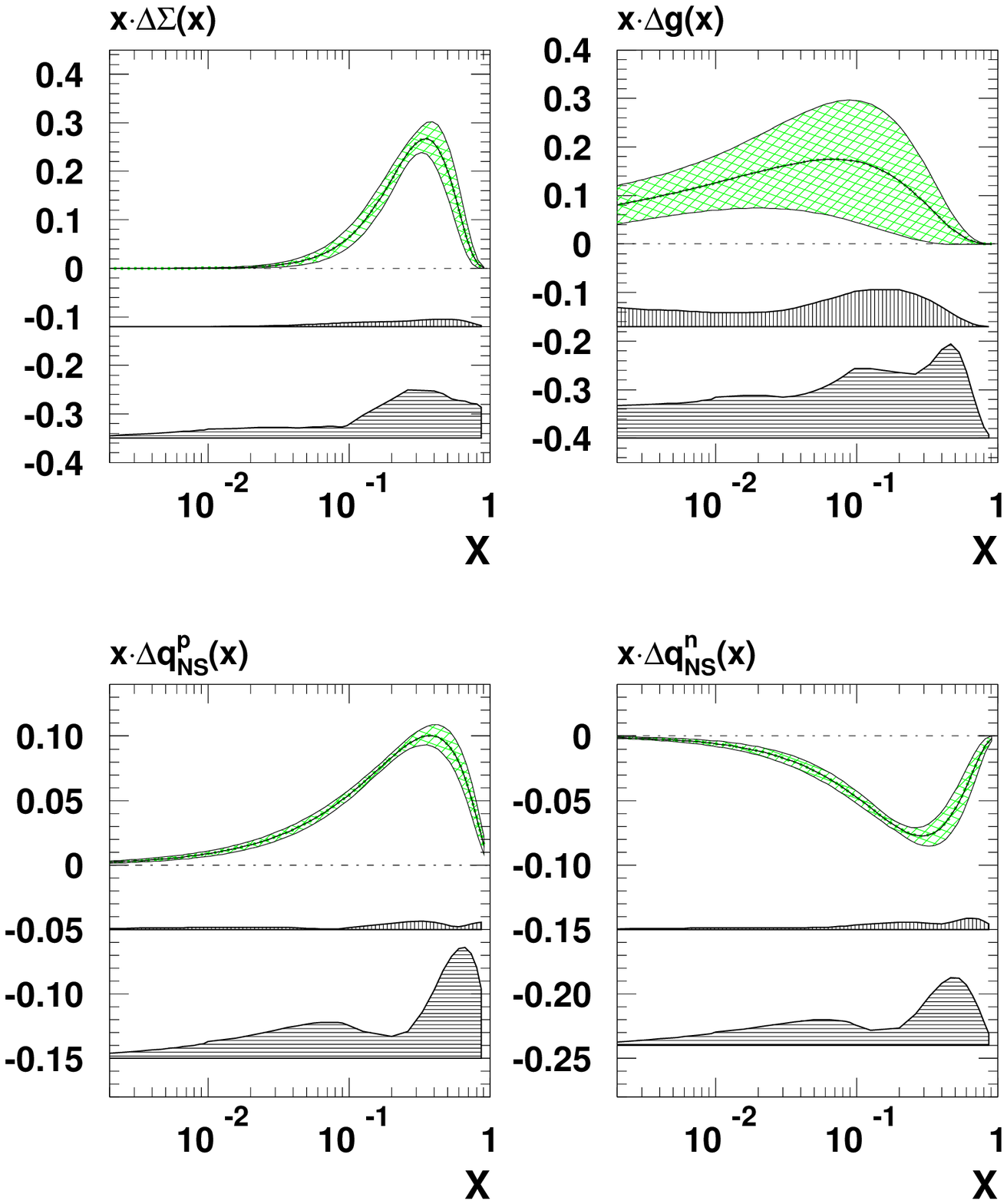}
\caption{Uncertainties on the parton distribution functions.
The cross-hatched band represents the statistical error only.
The experimental (top) and theoretical (bottom)
systematic errors are shown below the fitted distributions
\protect\cite{SMC_98b}.}
\label{fig:SMC_error}
\end{center}
\end{minipage}
\end{figure}

The difference of the two frequently used schemes, the $\overline{\mathrm MS}$ and the Adler--Bardeen (AB) scheme,
is best illustrated on the first moment of $g_1$
\begin{equation}
\Gamma_1(Q^2) = \int_0^\infty g_1(x,Q^2)\id x.
\end{equation}
While the non-singlet part of $\Gamma_1$ is scheme independent, the separation between gluon and quark
singlet contributions is as usual ambiguous beyond leading order. The singlet part of $\Gamma_1$ is 
often denoted as $1/9~a_0(Q^2) C^{\mathrm s}(Q^2)$ with the ``axial charge'' $a_0$ (see Eq.~\ref{eq:ejsr}).
In the  $\overline{\mathrm MS}$ scheme
there is no explicit gluon contribution to  $\Gamma_1$ ($C^{\mathrm g}(Q^2)\equiv 0$)
and thus $\Delta\Sigma(Q^2)=a_0(Q^2)$ is scale dependent. The AB scheme is tailored to define a scale independent
quark contribution, $\Delta\Sigma^{\mathrm AB}$, and to absorb the scale dependence in the anomalous gluon contribution
to $a_0$
\begin{equation}
\label{eq:anomal}
a_0(Q^2) = \Delta\Sigma^{\mathrm AB} - n_f \frac{\alpha_s}{2\pi}\Delta g(Q^2).
\end{equation}
The gluon distribution function is the same in the two scheme discussed above.
The quantity $\Delta\Sigma^{\mathrm AB}$ is often interpreted as the quark spin content of the nucleon. Therefore
a large value of $\Delta\Sigma^{\mathrm AB}$ requires a large positive gluon polarisation in order to be 
consistent with the small experimental value for $a_0$.

QCD analyses of the $g_1(x,Q^2)$ data were performed by several experimental \cite{E154_97b,SMC_98b} and theoretical
\cite{GeS_96a,GlR_95a,Str_99a,AlB_98a,LeS_99a} groups. A particular difficult problem
is the correct implementation of experimental and theoretical systematic errors. 
A detailed study of the propagation of these errors into the fitted parton distribution functions
was carried out by the SMC \cite{SMC_98b}. Fits in $x$ and $n$ space, AB and $\overline{\mathrm MS}$ schemes,
with different factorisation and renormalisation scales and different $Q_i^2$ were compared in this study. 
The $\overline{\mathrm MS}$ fit with $Q_i^2=1~\GeV^2$ to the proton, deuteron and neutron data is 
shown in Fig.~\ref{fig:SMC_qcd}.
The parton distribution functions obtained in the $\overline{\mathrm MS}$ and AB schemes agree well for
the non-singlet distributions (Fig.~\ref{fig:SMC_pdf}), while the $\Delta\Sigma(x)$ distributions must be
different for a non-vanishing gluon distribution. The negative part of $\Delta\Sigma(x)$ in the 
$\overline{\mathrm MS}$ scheme leads a smaller first moment than the positive distribution in the AB
scheme (see Eq.~\ref{eq:anomal}). The unexpected difference between the two gluon distributions 
reflects that the present data hardly constrain this quantity. For the
other distribution functions the theoretical uncertainty dominates (Fig.~\ref{fig:SMC_error}).

\begin{figure}[htb]
\begin{minipage}[t]{0.48\hsize}
\begin{center}
\epsfig{
bbllx=49, bblly=-20, bburx=397, bbury=543, 
width=0.85\hsize,file=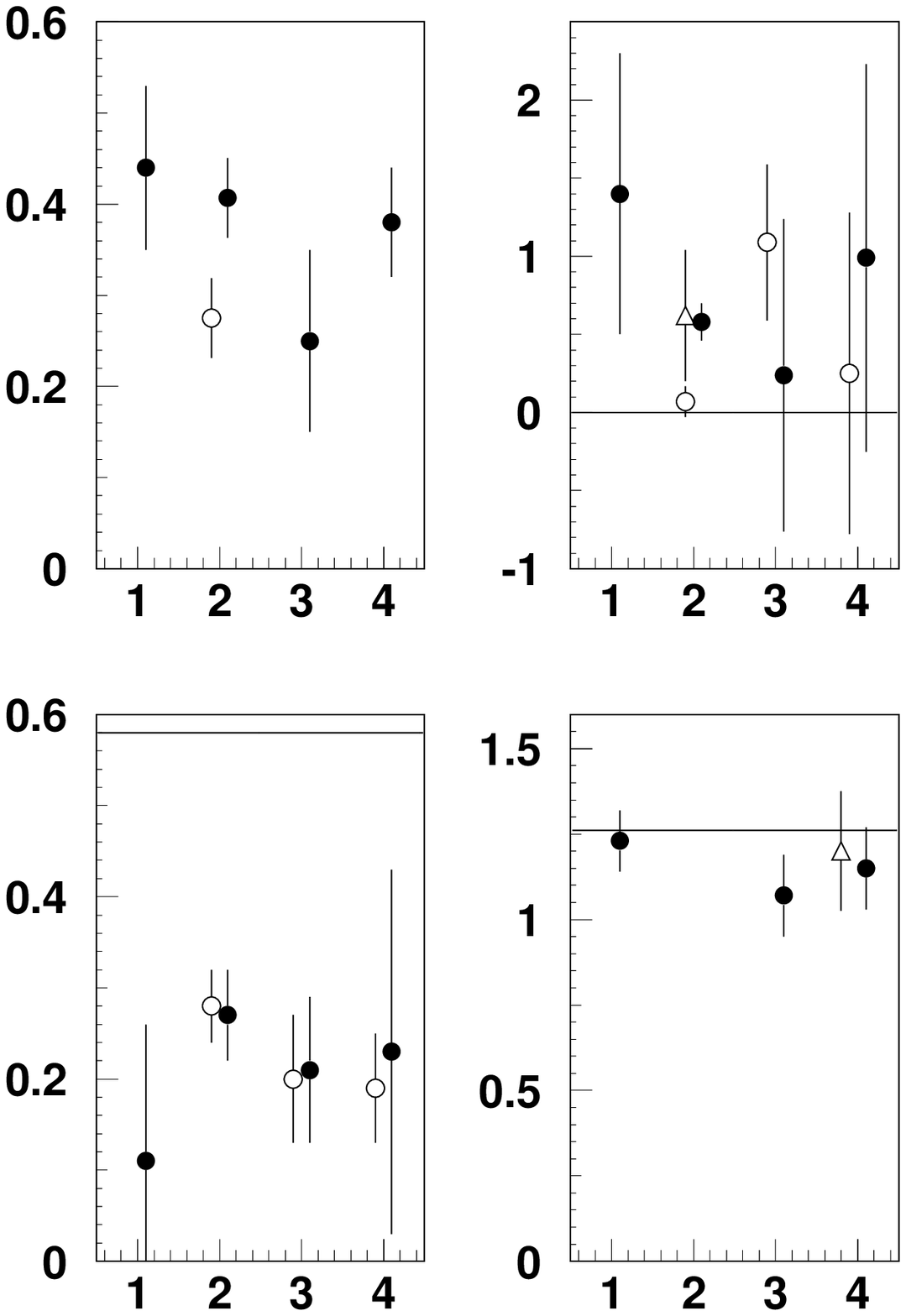}
\begin{picture}(0,0)(0,0)
\put(-170,300){a) \large$\Delta\Sigma$(AB)}
\put(-170, 10){c) \hskip 7mm \large$a_0$}
%\put(-170,  0){Ellis--Jaffe}
%\put(-170,-10){sum rule}
\put( -70,300){b) \hskip 5mm\large$\Delta$g}
\put( -70, 10){d) \hskip 1mm\large$|g_a/g_v|$}
%\put(  10, 70){Bjorken}
%\put(  10, 60){sum rule}
\end{picture}
\caption{Moments at $Q^2=1~\GeV^2$ from QCD fits
1--4 \protect\cite{AlB_98a,{LeS_99a},E154_97b,SMC_98b},
a)~``quark spin content'', b)~gluon polarisation,
c)~axial charge, Ellis--Jaffe sum rule, d)~axial coupling,
Bjorken sum rule.
}
\label{fig:moments}
\end{center}
\end{minipage}
\hfill
\begin{minipage}[t]{0.48\hsize}
\begin{center}
\epsfig{width=\hsize,file=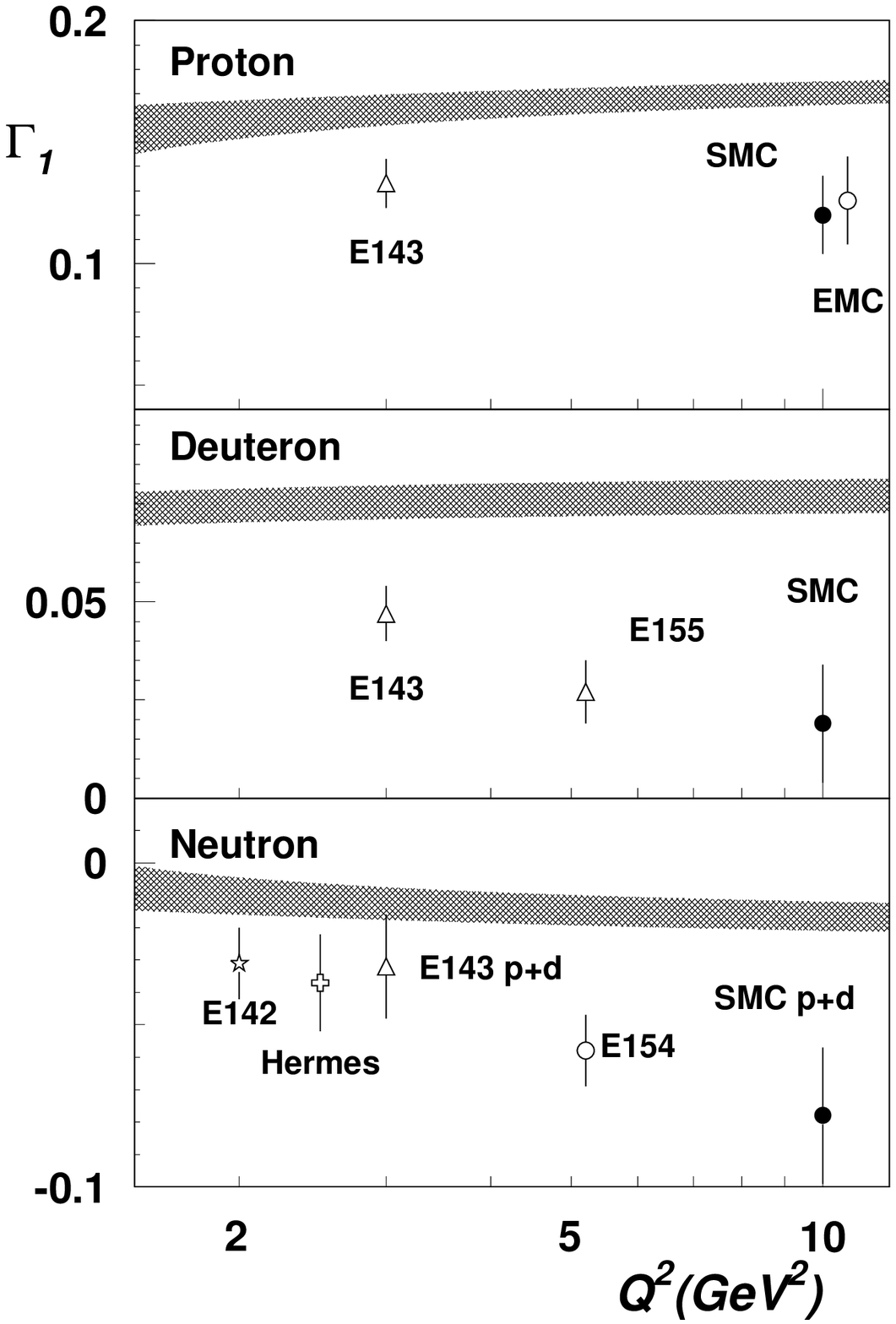}
\caption{First moments of $g_1$ for the proton, the deuteron and the
neutron at the typical $Q^2$ of the data. The hatched bands show the
prediction of the Ellis-Jaffe sum rule.
}
\label{fig:ejall}
\end{center}
\end{minipage}
\end{figure}

\section{Sum rules and moments}

From the QCD analyses the first moments and the axial charge
can be determined directly by integration of the appropriate
parton distribution functions. Figure~\ref{fig:moments} summarises
the results from fits by (1) Altarelli, Ball, Forte, Ridolfi, Fit~B
\cite{AlB_98a}, (2) Leader, Sidorov and Stamenov \cite{LeS_99a},
(3) E154 \cite{E154_97b} and (4) the SMC \cite{SMC_98b}. All fits
use $Q^2_i=1~\GeV^2$, except the E154 fit which uses 
$Q^2_i=0.34~\GeV^2$.

The results for $\Delta g(1~\GeV^2)$ vary strongly, however all 
fits give positive values. Fit (2) only takes into account the statistical
errors. A slight change of the parametrisation of the gluon
distribution for this fit results in a quite different value
for $\Delta g$ (open triangle, Fig.~\ref{fig:moments}b). 
Often the violation of the Ellis-Jaffe sum rule is attributed to
the axial anomaly and it is suggested that instead of the axial
charge, $a_0$, the scale invariant quark spin content,
$\Delta\Sigma^{\mathrm AB}$, should be consistent
with the Ellis-Jaffe expectation of about 0.6. This is not 
supported by the data (Fig.~\ref{fig:moments}a), which favour a
value around 0.4. 
In some fits the axial coupling constant $|g_a/g_v|$ 
was treated as a free parameter.
The agreement with the experimental value of 1.2601$\pm0.0025$ and thus
with the Bjorken sum rule \cite{Bjo66} is
demonstrated in Fig.~\ref{fig:moments}d. The moments obtained
from the next-to-leading order fits take only correction to
order ${\cal O}(\alpha_s^2)$ into account, while the corrections
to the Ellis--Jaffe and Bjorken sum rules are known up to order 
${\cal O}(\alpha_s^3)$ \cite{Lar_97a}.

A direct evaluation of the first moment $\Gamma_1$ from the $g_1$
data requires the evolution of the data to a common value of $Q^2$ in
the measured range and extrapolations to $x=0$ and $x=1$. While
the latter is unproblematic, the small-$x$ extrapolation has long been
debated in the literature. The presently most reliable procedure
is to use the QCD fits for $x\rightarrow0$.
The present status of the world data for $\Gamma_1$ for the proton,
the deuteron and the neutron is shown in Fig.~\ref{fig:ejall}.
All experiments show a violation of the Ellis-Jaffe sum rule 
independent of the target.
The three analyses using a QCD fit for the extrapolation to $x=0$
(SMC, E154, E155) show the strongest deviation due to the negative 
contribution from this extrapolation. This is most obvious for the
neutron, where Regge-type extrapolations (E142, E143, Hermes)
lead to first moments almost compatible with the Ellis-Jaffe
value.

Assuming SU(3) flavour symmetry the first moment of $g_1$ is given
by
\begin{equation}
\Gamma_1(Q^2) = C^{\mathrm s}(Q^2)a_0(Q^2) 
+ \frac{1}{12}\left(\left|\frac{g_a}{g_v}\right|-\frac{1}{3}(3F-D)\right)
C^{\mathrm ns}(Q^2).
\label{eq:ejsr}
\end{equation}
%The SMC, E154, and E155 have evaluated the first moments using QCD 
%extrapolations.
In order ${\cal O}(\alpha_s^3)$ one obtains
from the first moments the following values for the axial charge, $a_0$:
$0.12\pm0.15$ (SMC proton),  $0.06\pm0.13$ (SMC deuteron), $0.18\pm0.10$
(E154 neutron) and  $0.14\pm0.07$ (E155 deuteron). The precise E155 data  
dominate the average of $a_0=0.14\pm0.05$, which is somewhat lower
than the value obtained directly from the QCD fits $a_0\simeq0.20$ 
(Fig.~\ref{fig:moments}c).
The agreement is satisfactory, in particular in view of the many differences
in the procedures which include: the order in $\alpha_s$, SU(3) symmetry,
fitting of $|g_a/g_v|$. For the evaluation of $a_0$ only those values of
$\Gamma_1$ were considered, in whose evaluation a QCD extrapolation to $x=0$
was used.

\begin{figure}[htb]
\begin{minipage}[t]{0.47\hsize}
\begin{center}
\epsfig{width=\hsize,file=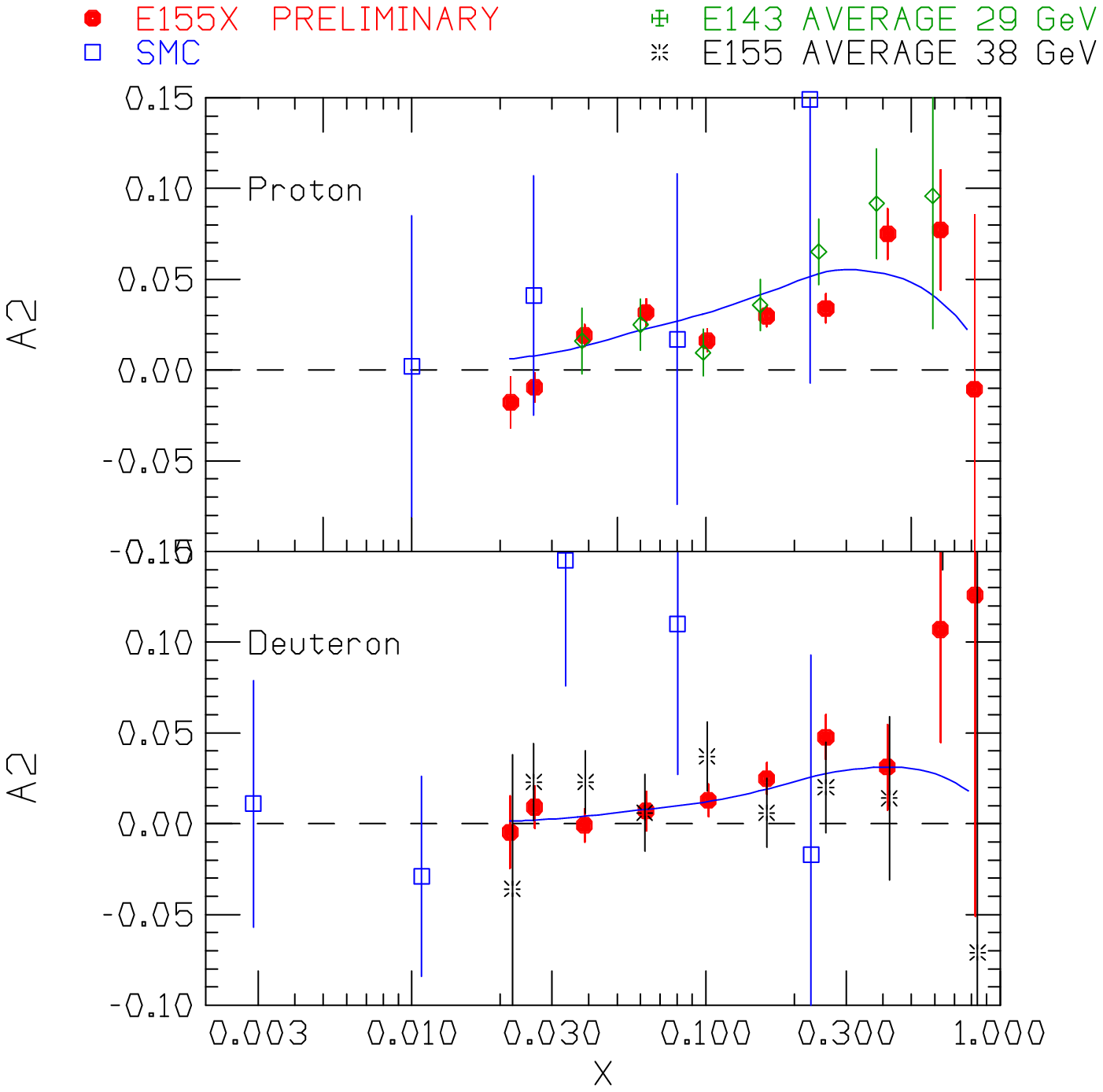}
\caption{Preliminary $A_2$ asymmetries from the E155x experiment for
the proton (top) and the deuteron (bottom) \protect\cite{steve}.
The solid line indicates the $A_2^{\mathrm WW}$ term.}
\label{fig:E155x_A2}
\end{center}
\end{minipage}
%\end{figure}
\hfill
%\begin{figure}[htb]
\begin{minipage}[t]{0.47\hsize}
\begin{center}
\epsfig{width=\hsize,file=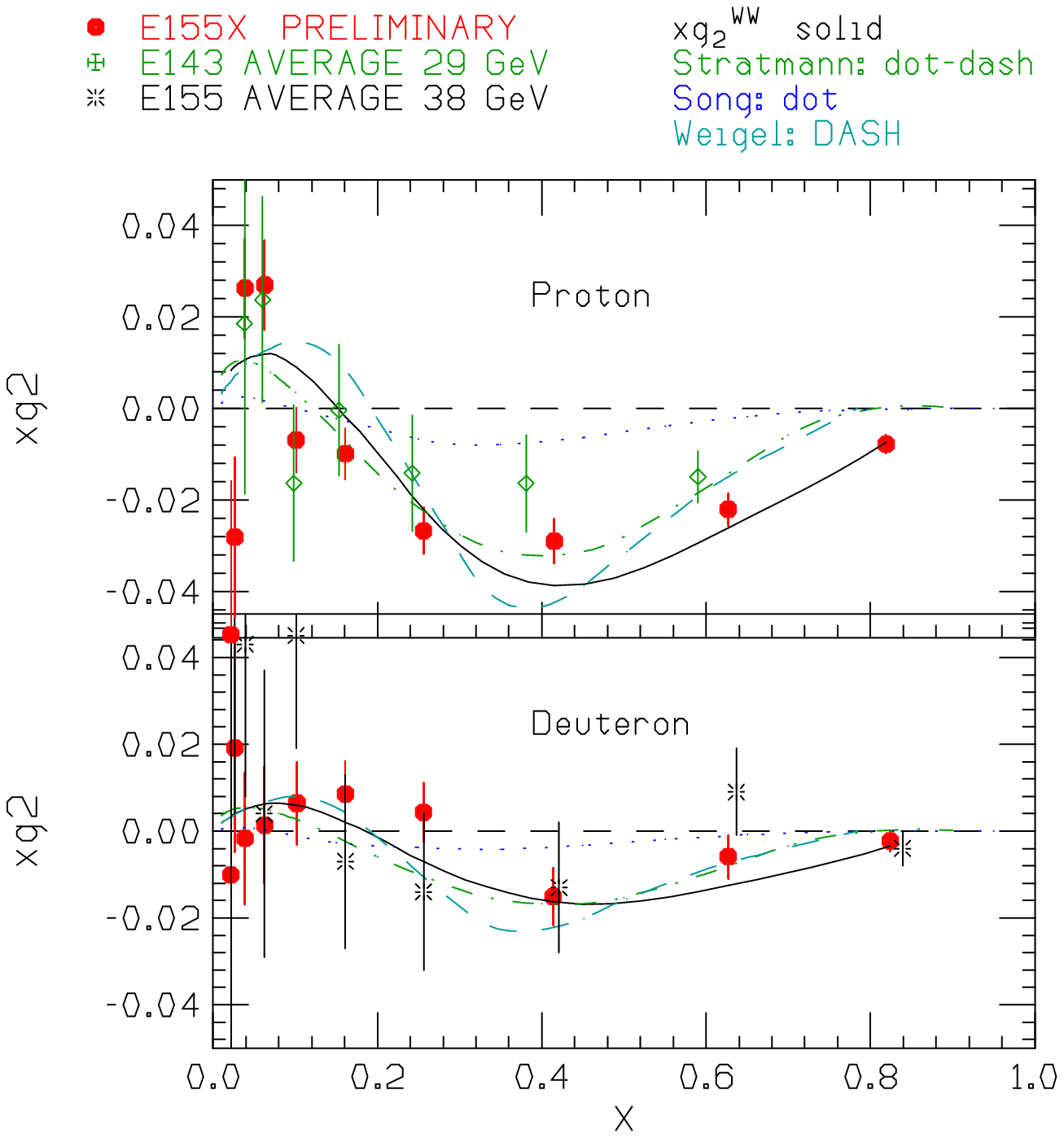}
\caption{Preliminary E155x $xg_2(x)$ data for
the proton (top) and the deuteron (bottom) \protect\cite{steve}.
Also shown are
the  $g_2^{\mathrm WW}$ term (solid) and several bag model
calculations (dashed and dotted).}
\label{fig:E155x_xg2}
\end{center}
\end{minipage}
\end{figure}

\section{Status of $A_2$ measurements}
The structure function $g_2$ offers the unique possibility to study a pure
twist-3 effect.  It can be split into a twist-2 term, $g_2^{\mathrm WW}$ \cite{WaW77}, which is 
calculable from $g_1$ and the twist-3 term $\overbar g_2$
\begin{eqnarray}
{g_2}(x,Q^2)&=&{g_2^{\rm WW}}+{\bar g_2}(x,Q^2),\\
{g_2^{\rm WW}}(x,Q^2)&=&-g_1(x,Q^2)+\int_x^1\frac{g_1(y,Q^2)}{y}\,\id y.
\end{eqnarray}
Measurements by the SMC \cite{SMC_94a} and by the E142 \cite{E142_96a}, E143 \cite{E143_98a},
E154 \cite{E154_97c}
and E155 \cite{Mitchell_99} experiments showed that the
corresponding asymmetry, $A_2$, is much smaller than its positivity limit
$|A_2|\le\sqrt{R}$.
The very precise preliminary data from the dedicated $g_2$ experiment E155x \cite{steve, E155_web} now
clearly establishes a non-vanishing $g_2$ for both, the proton and the deuteron.
The data are in line with the Wandzura-Wilczek term, $g_2^{\mathrm WW}$, and any
possible twist-3 term must be small ({Figs.~\ref{fig:E155x_A2} and
~\ref{fig:E155x_xg2}).

\begin{figure}[htb]
\begin{center}
\epsfig{width=0.5\hsize,file=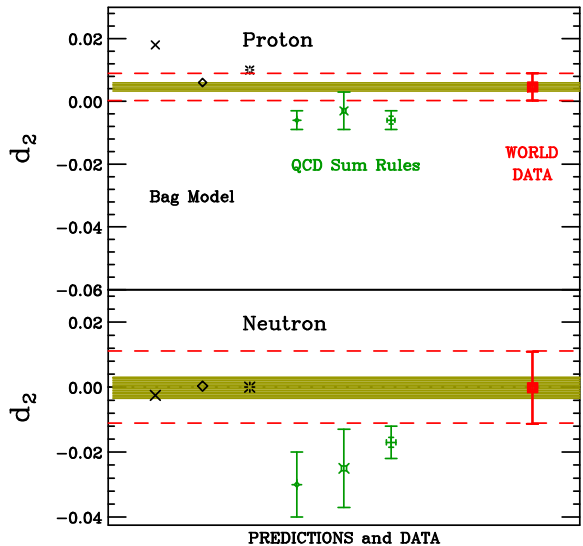}
\caption{The twist-3 matrix element $d_2$ of the proton and the neutron.
The dark band indicates the expected uncertainty of the E155x data, the
values are not yet released.
Also shown are bag model and QCD sum rule calculations \protect\cite{steve}.}
\label{fig:d2}
\end{center}
\end{figure}

The twist-3 matrix element $d_2$
\begin{equation}
d_2 = 3 \int_0^1 x^2 \,{\bar g_2}(x,Q^2)\,\id x
\end{equation}
is related to higher twist effects for the first moments and thus
for the Bjorken sum rule.
The world data are
$d_2^{\mathrm p}=0.007\pm0.004$ and  $d_2^{\mathrm p}=0.004\pm0.010$ for the
proton and the neutron, respectively. All data come from the SLAC experiments
E142--E155 and are shown in Fig.~\ref{fig:d2}. The present precision is not
sufficient to decide between bag model and QCD sum rule calculations. 
However, the precision expected for the E155x data might allow us to
distinguish between the two classes of predictions.

\begin{figure}[htb]
\begin{minipage}[t]{0.54\hsize}
\begin{center}
\epsfig{width=\hsize,file=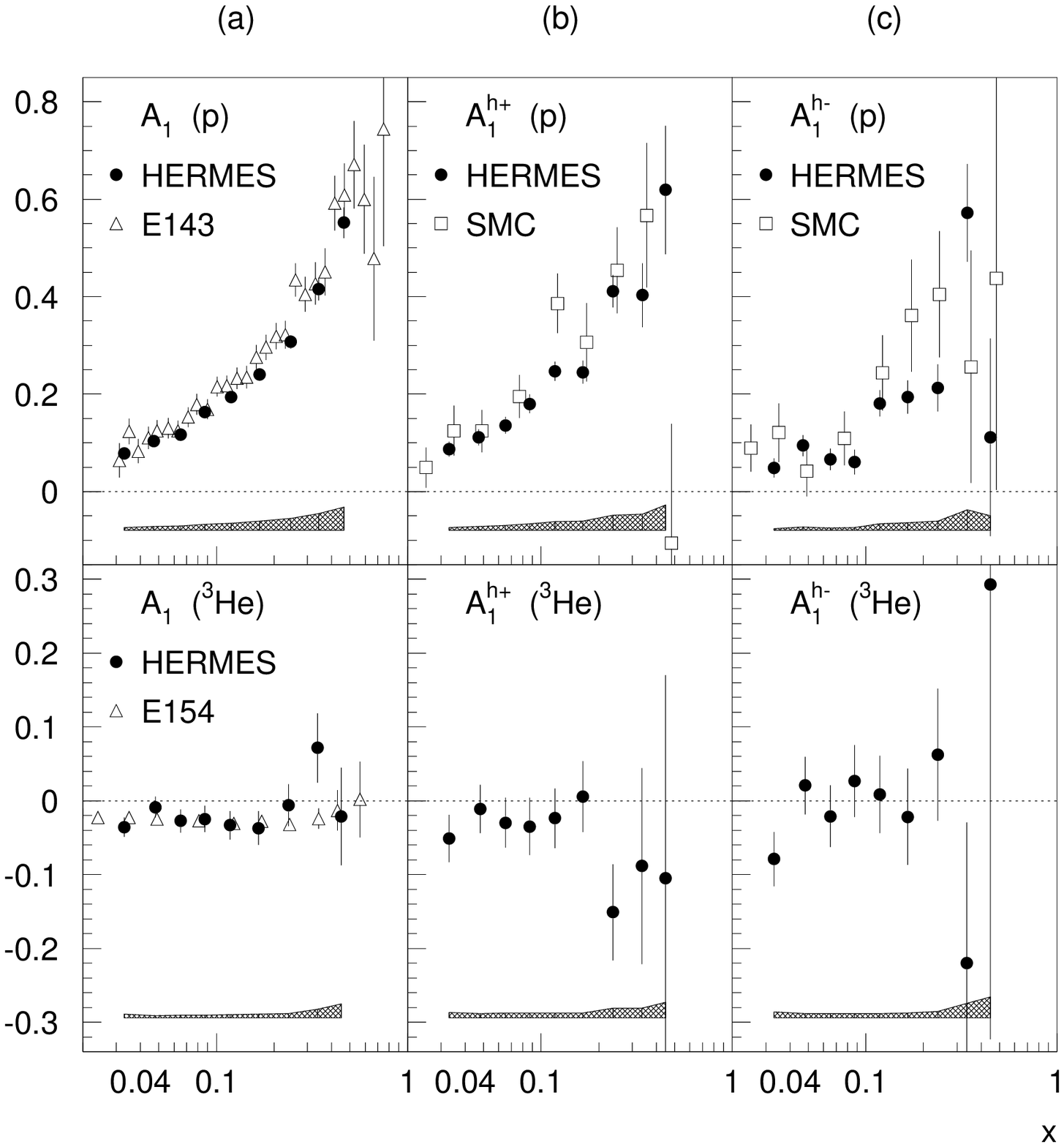}
\caption{Inclusive (a) and semi-inclusive Asymmetries of positive (b) and
negative (c) hadrons for the proton (top) and $^3$He (bottom) \cite{Hermes_99a}.}
\label{fig:A_semi}
\end{center}
\end{minipage}
%\end{figure}
\hfill
%\begin{figure}[htb]
\begin{minipage}[t]{0.42\hsize}
\begin{center}
\epsfig{width=\hsize,file=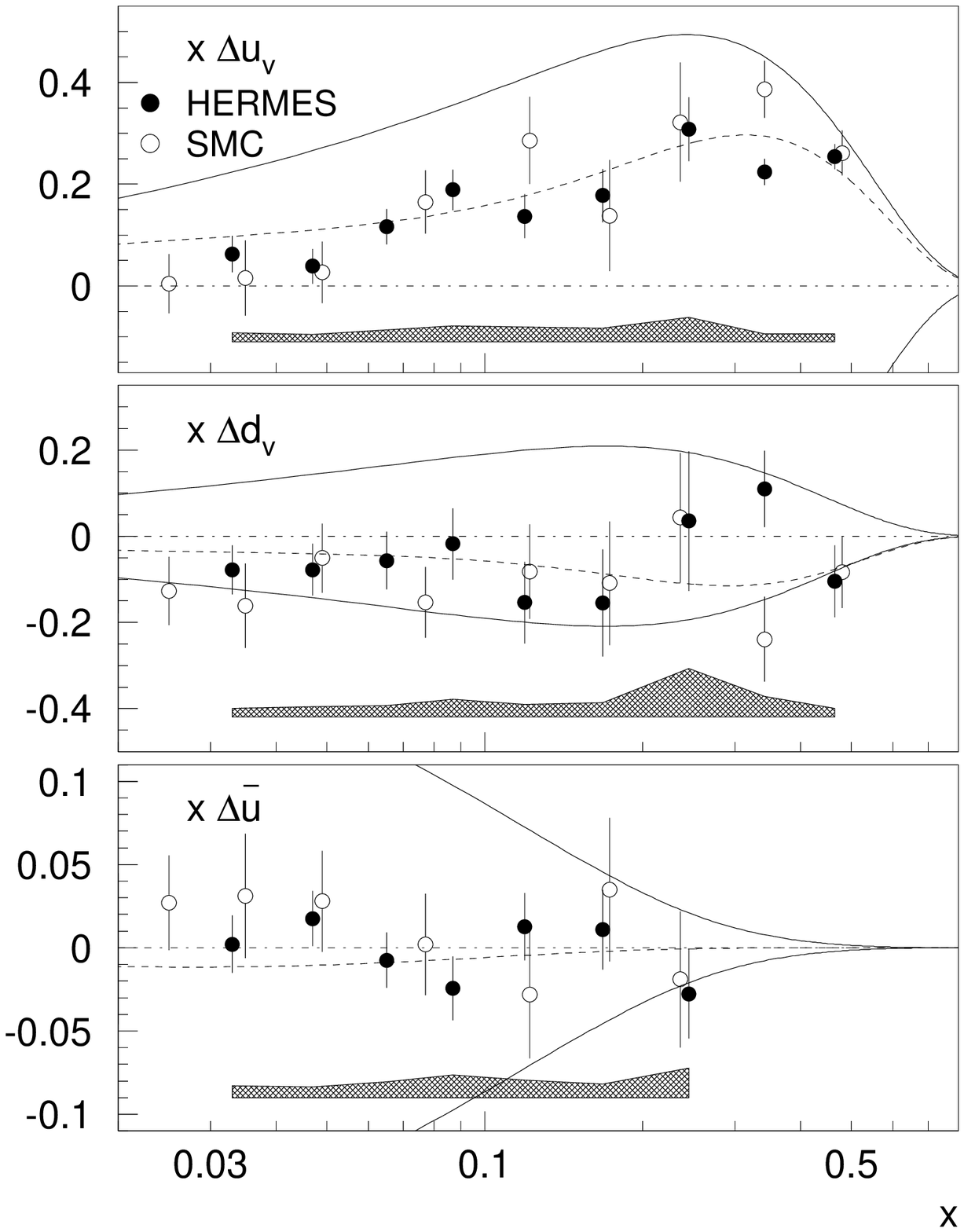}
\caption{The valence and sea quark distributions at $Q^2=2.5~\GeV^2$
from the SMC and Hermes \cite{Hermes_99a}.  Also shown are the
positivity limit (solid) and a parametrisation from \cite{GeS_96a} (dashed).}
\label{fig:q_semi}
\end{center}
\end{minipage}
\end{figure}

\section{Semi-inclusive data}
In semi-inclusive processes a hadron is detected in addition to the
scattered lepton. Due to the favoured fragmentation of e.g.\ a u-quark
(d-quark) into $\pi^+$ ($\pi^-$) one can perform a flavour separation
in the parton distribution functions. The asymmetries from
the SMC and Hermes experiments \cite{SMC_97a,Hermes_99a} for positive
and negative hadrons are shown in Fig.~\ref{fig:A_semi}. A full particle
identification was not available in these experiments. From these asymmetries
one can derive the polarised valence parton distributions, $\Delta u_{\mathrm v}$
and $\Delta d_{\mathrm v}$, as well as the anti-quark distribution. For the
latter in the Hermes analysis it was assumed that
\begin{equation}
\frac{\Delta \bar u(x)}{\bar u(x)}=
\frac{\Delta \bar d(x)}{\bar d(x)}=
\frac{\Delta \bar s(x)}{\bar s(x)}.
\end{equation}
The results are shown in Fig.~\ref{fig:q_semi} for the Hermes $x$ range,
while the SMC extent to $x=3\cdot10^{-5}$. The Hermes systematic error
is shown by the band, while that for the other data is included in the
error bar.

Recently first results were reported involving azimuthal asymmetries.
The SMC analysed data taken with transverse target polarisation in terms of
the Collins angle and found an asymmetry of $A_N=0.11\pm0.06$ and
$-0.02\pm0.06$ for positive and negative hadrons, respectively \cite{Bra_99a}.
Hermes reported an azimuthal asymmetry of $A_N=0.022\pm0.005\pm0.003$ for $\pi^+$ and 
$-0.002\pm0.006\pm0.004$ for $\pi^-$ using longitudinal target
polarisation \cite{Hermes_99b}. These data represent the first glimpse at the
chiral-odd transversity structure function, $h_1$.

Hermes has studied asymmetries in events with high-$p_T$
oppositely-charged hadron pairs \cite{Hermes_99c}. 
The asymmetry is insignificant for events with
positive hadrons with $p_T>1.5~\GeV^2$ and events where both hadrons 
have $p_T<1~\GeV^2$. For $p_T^{\mathrm h^+}>1~\GeV^2$ and
$p_T^{\mathrm h^-}>1.5~\GeV^2$ there are two data points showing a
non-zero asymmetry. The asymmetry in this region is $-0.28\pm0.12\pm
0.02$. This has been interpreted in terms of a positive gluon 
polarisation.

\section{Summary and Outlook}
A large amount of high quality data is now available for the structure
functions $g_1$ in a large range of $x$ and $Q^2$. The data are
consistent with QCD evolution and the NLO fits taught us much about
the polarised parton distribution functions. In contrast to the unpolarised
case were neutrino scattering provides information on the flavour
separation, in polarised deep inelastic scattering  we must rely on the 
semi-inclusive processes using flavour tagging by hadrons. 
Semi-inclusive processes are also the
only way in DIS to observe the chiral-odd structure function $h_1$, for
which first information was presented this year. Big progress has been
made in the knowledge of $g_2$.

In spite of all the precise data the spin puzzle is not yet resolved
and a direct measurement of the gluon polarisation and if possible of
the orbital angular momentum is inevitable. New experiments will study
this quantity at CERN (COMPASS) and RHIC starting data taking 2001.
An exciting longer-term perspective is the possibility of a polarised 
proton beam at Hera, which would open up a new window to the nucleon's 
spin structure.

\bigskip
I appreciate the help of E.~Rondio and S.~Rock and my colleagues from
the SMC and COMPASS collaborations in the preparation of this talk.
The organizers of LP99 I thank for the invitation to this
inspiring conference.

\end{document}